\documentclass{article}
\usepackage{amsmath,amssymb,cite,epsfig,graphicx,fancyhdr}
\begin{document}
\lhead{Wronskian differential formula for confluent SUSY QM}
\rhead{Bermudez, Fern\'andez, Fern\'andez-Garc\'ia	}

\title{Wronskian differential formula for confluent\\ supersymmetric quantum mechanics}
\author{David Bermudez\footnote{{\it email:} dbermudez@fis.cinvestav.mx},\, David J. Fern\'andez C.\footnote{{\it email:} david@fis.cinvestav.mx} \\ \vspace{2mm}
{\sl Departamento de F\'{\i}sica, Cinvestav, A.P. 14-740, 07000 M\'exico D.F., Mexico}\\
Nicol\'as Fern\'andez-Garc\'ia\footnote{{\it email:} jfernandezg@ipn.mx}\\
{\sl Secci\'on de Estudios de Posgrado e Investigaci\'on, UPIITA, IPN, Av. IPN 2508,}\\ {\sl 07340 M\'exico D.F., Mexico}}

\date{}

\maketitle

\begin{abstract}
A Wronskian differential formula, useful for applying the confluent second-order SUSY transformations to arbitrary potentials, will be obtained. This expression involves a parametric derivative with respect to the factorization energy which, in many cases, is simpler for calculations than the previously found integral equation. This alternative mechanism shall be applied to the free particle and the single-gap Lam\'e potential.
\end{abstract}


\section{Introduction}

Over time, there has been much interest in the study and generation of exactly solvable potentials in quantum mechanics. One of the simplest techniques to carry out these purposes is the supersymmetric quantum mechanics (SUSY QM), which is based on the intertwining relationship. In fact, this technique together with the factorization method and Darboux transformation are equivalent procedures for generating solvable Hamiltonians from a given initial one \cite{MS91,BS97,MRO04,FF05,Fer10}. In the simplest case of this approach both Hamiltonians are related by a first-order differential operator. However, the fact that the zeros of the transformation function are mapped into singularities of the new potential imposes certain restrictions on this method.

An alternative to avoid this difficulty is to employ the SUSY QM of higher order (for more details and applications see \cite{MRO04,FF05,Fer10,AIS93,AICP95,FGN98,MNR000,AC04,PN10,PAN11}). The most elementary version of such a generalization is the confluent second-order SUSY QM \cite{MNR000,FS03}, for which the two involved factorization energies converge to the same value. However, the main problem for implementing this method has to do with the difficulty to calculate certain integrals which arise in the treatment.

On the other hand, it has been shown \cite{FS05} that in the confluent case the Wronskian formula is preserved if solutions closing a Jordan chain of length two are used as seeds for implementing the algorithm. In this article we will take advantage of this fact by introducing a differential version of the technique which will preserve as well the general Wronskian formula and will avoid to evaluate the previously mentioned integrals. In this way, an alternative calculational tool will be available for implementing the confluent second-order SUSY QM.

The paper is organized as follows: in Section 2 we shall review the standard approach to the second-order SUSY QM for the confluent and non-confluent cases. A differential version of the Wronskian formula for the confluent case will be presented in Section 3. In Section 4 this alternative method will be applied to the free particle and the single-gap Lam\'e potentials. A summary of our original results and some conclusions are contained in Section 5.

\section{Second-order SUSY QM}

In the second-order SUSY QM one typically starts from the following intertwining relationship \cite{MRO04,Fer10,AIS93}
\begin{equation}
\widetilde{H} B^{+} = B^{+} H,\label{2S_Entrelaza}
\end{equation}
where
\begin{align}
H &=-\frac{d^2}{d x^2} + V(x),\label{2S_H}\\
\widetilde{H} &= - \frac{d^2}{d x^2} + \widetilde{V}(x),\label{2S_Ht}\\
B^{+} &= \frac{d^2}{d x^2} + g(x)\frac{d}{dx}+h(x).\label{2S_Bmas}
\end{align}
The aim is to determine $\widetilde{V}(x)$, $g(x)$ and $h(x)$ supposing that the initial potential $V(x)$ is known. The solution to this problem is given by
\begin{align}
\widetilde{V}(x) & = V + \, 2g',\label{eqv}\\
h(x) & = -\frac{g'}{2} +\frac{g^2}{2} - V + d,
\end{align}
where $g(x)$ must fulfil the non-linear differential equation
\begin{equation}
\frac{gg''}{2}-\frac{g'^2}{4}+g^2\left(-g'+\frac{g^2}{4}-V+d \right)+c=0, \label{eqg}
\end{equation}
and $c,d\in\mathbb{R}$ are two integration constants. In order to solve \eqref{eqg} it is used the ansatz \cite{FGN98}
\begin{equation}
g'(x) = g^2+2\gamma g - 2 \xi, \label{ans}
\end{equation}
where $\gamma(x)$ and $\xi(x)$ are functions to be determined. Substituting \eqref{ans} in Equation \eqref{eqg}, it turns out that $\xi^2=c$ and the following Ricatti equation must be satisfied
\begin{equation}
\gamma' + \gamma^2 = V - \epsilon,\label{2S_Ricatti}
\end{equation}
with $\epsilon = d + \xi$. Thus, the initial problem defined by Equations (\ref{2S_Entrelaza}-\ref{2S_Bmas}) has been reduced to find the function $\gamma$ and, consequently, to solve Equation (\ref{2S_Ricatti}). This Riccati equation can be linearized by using $\gamma=u'/u$, which leads to the following stationary Schr\"odinger equation for $H$:
\begin{equation}
Hu = -u''+Vu=\epsilon u. \label{sch}
\end{equation}
The kind of seed solution $u$ employed for constructing the transformation depends on the factorization energy $\epsilon$ and, consequently, on the sign of $c$. For $c\neq 0$ one gets the so called real and complex cases while for $c=0$ the confluent one is obtained.

\subsection{Non-confluent case ($c \neq 0$)}

Let us denote the two different factorization energies by $\epsilon_1 \equiv d+c^{1/2}$ and $\epsilon_2 \equiv d-c^{1/2}$, which includes both, the real and complex cases. Note that the ansatz \eqref{ans} indeed gives place to two equations
\begin{align}
g'&=g^2+2\gamma_1 g - (\epsilon_1 - \epsilon_2),\\
g'&=g^2+2\gamma_2 g - (\epsilon_2 - \epsilon_1).
\end{align}
By subtracting both it is obtained
\begin{equation}
g=\frac{\epsilon_1-\epsilon_2}{\gamma_1-\gamma_2}=-\{\ln[W(u_1,u_2)]\}', \label{solg}
\end{equation}
where $W(f,h)=fh'-f'h$ is the Wronskian of $f$ and $h$. The Wronskian in \eqref{solg} should not have zeros in order to avoid singularities in $g$ and, consequently, in $\widetilde{V}$. Substituting \eqref{solg} in \eqref{eqv} one gets
\begin{equation}
\widetilde{V}=V-2\{\ln[W(u_1,u_2)]\}''.\label{newV}
\end{equation}
This expression has been used to construct a wide variety of potentials $\widetilde V$ departing from a given initial one $V$ by choosing two appropriate solutions of \eqref{sch}. In fact, this treatment is so versatile that it can be implemented by taking as seeds even the Gamow vectors, which gives place to complex potentials with known spectra (see e.g. \cite{FR1}).

\subsection{Confluent case ($c = 0$)}

In the confluent case ($c=0$) both energies converge to just one, $\epsilon_2 \rightarrow \epsilon_1$, and the ansatz of Equation \eqref{ans} becomes
\begin{equation}
g' =g^2+2\gamma_1 g.
\end{equation}
This Bernoulli equation has a general solution given by
\begin{equation}
g(x)=-\{\ln[w(x)]\}',
\end{equation}
where
\begin{equation}\label{wconfluente}
w(x)=w_0-\int_{x_0}^{x}u_1^2(y)dy,
\end{equation}
with $w_0$ and $x_0$ being real constants which can be chosen at will in order to avoid singularities in $g(x)$.

On the other hand, let us consider the following pair of generalized eigenfunctions of $H$, of first and second rank, associated to $\epsilon_1$ \cite{FS05,DK67,EJM06},
\begin{align}
(H-\epsilon_1)u_1 & = 0,\label{uv1}\\
(H-\epsilon_1)u_2 & = u_1, \label{uv2}
\end{align}
which is known as Jordan chain of length two. By solving Equation \eqref{uv2} for $u_2$ through the method of variation of parameters, supposing that $u_1$ is given, we get
\begin{equation}
u_2=\left(k+\int \frac{w(x)}{u_1^2(x)}dx\right)u_1(x).\label{u2}
\end{equation}
Moreover, by using the following Wronskian identity
\begin{equation}
W(f,hf)=h'f^2,\label{wrons}
\end{equation}
which is valid for two differentiable arbitrary functions $f$ and $h$, it is straightforward to show that
\begin{equation}
w(x)=W(u_1,u_2).\label{wW}
\end{equation}
Therefore, the Wronskian formula of the non-confluent second-order SUSY QM given by Equation \eqref{newV} is preserved for the confluent case \cite{FS05}. Moreover, it can be used to construct a one-parameter family of exactly solvable potentials for each solution $u_1$ of the initial stationary Schr\"odinger equation associated to $\epsilon_1$. However, if $u_1$ has an involved explicit form the task of evaluating the corresponding integrals is not simple. In the next section we shall present an alternative version of the Wronskian formula for the confluent case which will make unnecessary the evaluation of the integrals of Equations \eqref{wconfluente} and \eqref{u2}.

\section{Wronskian differential formula for the confluent SUSY QM}

Let us look for now the general solution of Equation \eqref{uv2} in a slightly different way. Let $u_1$ denote once again the given solution of \eqref{uv1}. It is well known that the general solution of the inhomogeneous second-order differential equation \eqref{uv2} takes the form:
\begin{equation}
u_2=u_2^{h}+u_2^{p},
\end{equation}
where $u_2^{h}$ is the general solution of the homogeneous equation and $u_2^{p}$ denotes a particular solution of the inhomogeneous one. Since the homogeneous equation is of second order, it has two linearly independent solutions. They can be taken as $u_1$ and its orthogonal function $u_1^\perp$ defined by $W(u_1,u_1^\perp) = 1$. The last equation can be immediately solved for $u_1^\perp$, yielding
\begin{equation}
u_1^\perp (x) = u_1(x)\int\frac{dx}{u_1^2(x)}.
\end{equation}
Then, it turns out that
\begin{equation}
u_2^{h}=Cu_1+Du_1^\perp,
\end{equation}
with $C,\,D \in \mathbb{R}$.

In order to find the particular solution $u_2^{p}$, let us suppose from now on that $u_1$ and its parametric derivative with respect to $\epsilon_1$, $\frac{\partial u_1}{\partial\epsilon_1}$, are well defined continuous functions in a neighbourhood of $\epsilon_1$. Hence, by deriving Equation \eqref{uv1} with respect to $\epsilon_1$ it is obtained:
\begin{equation}
\left(H-\epsilon_1\right)\frac{\partial u_1}{\partial\epsilon_1} = u_1, \label{part}
\end{equation}
where the partial derivatives of $u_1$ with respect to $\epsilon_1$ and $x$ have been interchanged. It should be clear now that (compare Equations \eqref{uv2} and \eqref{part})
\begin{equation}
u_2^{p} = \frac{\partial u_1}{\partial\epsilon_1}
\end{equation}
is the particular solution of the inhomogeneous equation we were looking for. Finally, the general solution of equation \eqref{uv2} is given by
\begin{equation}
u_2= C u_1 + D u_1^\perp + \frac{\partial u_1}{\partial\epsilon_1} .\label{u2a}
\end{equation}
From this equation we can easily calculate the Wronskian of the two solutions of the Jordan chain as
\begin{equation}
W(u_1,u_2)= D + W\left(u_1,\frac{\partial u_1}{\partial\epsilon_1}\right).
\end{equation}
Thus, the general Wronskian formula of Equation \eqref{newV} becomes now
\begin{equation}
\widetilde{V}=V-2\left\{\ln\left[D + W\left(u_1,\frac{\partial u_1}{\partial\epsilon_1}\right)\right]\right\}'', \label{Dv}
\end{equation}
which represents an alternative way to calculate the new potential $\widetilde{V}$ through the confluent second-order SUSY transformation.

Note that a special case of equation \eqref{Dv} has been addressed previously, for $D=0$ and the free particle potential \cite{Mat92,Sta95}. In these works, the particular solution $\frac{\partial u_1}{\partial\epsilon_1}$ was taken directly as the seed solution $u_2$ and thus the constant $D$, which arises from the non-trivial term involving the orthogonal function $u_1^\perp$ (see the second term of the right hand side of Equation \eqref{u2a}), never appears in those treatments.

An additional point is worth to remark: without the constant $D$ the confluent second-order SUSY partner potential $\widetilde V$ will often have singularities. The freedom we have here for choosing this constant endows us with the possibility to generate families of non-singular potentials for a wide set of factorization energies.

\section{Applications}

We are going to use Equation \eqref{Dv} now to implement a confluent second-order SUSY transformation for two simple systems. The first of them is the free particle, where both the differential and the integral versions of the confluent SUSY QM are easily applicable since the derivatives and the integrals involved are not difficult to calculate. The second one is the single-gap Lam\'e potential, for which the previously found integral Equation \eqref{wconfluente} is not easy to apply, since the integrals of elliptic functions are complicated to evaluate. As far as we know, the confluent second-order SUSY transformation has been never applied before to this potential.

\subsection{Free particle}\label{secfree}

The free particle is not subject to any force so that the corresponding potential is constant; without loss of generality, let us take $V(x)=0$. In order to obtain non-singular confluent second-order SUSY partner potentials one has to use as transformation function, in general, a solution $u_1$ to the stationary Schr\"odinger equation \eqref{sch} such that $W(u_1,u_2)\neq 0 \ \forall \ x\in{\mathbb R}$. This is achieved by demanding that $u_1$ vanishes at one of the boundaries of the $x$-domain (see \cite{FS03,FS05}). In particular, for the free particle these solutions are $\{e^{\kappa_1 x},e^{-\kappa_1 x}\}$ with the condition that $\kappa_1$ and $\epsilon_1$ satisfy the `dispersion relation' $\epsilon_1 = -\kappa_1^2, \ \kappa_1>0$.

We are going to use one of these solutions to perform the SUSY transformation, e.g., $u_1 = e^{\kappa_1 x}$; the other case can be obtained through a spatial reflection. Thus, the parametric derivative can be calculated using the chain rule as
\begin{equation}
\frac{\partial u_1}{\partial\epsilon_1} = \frac{d\kappa_1}{d\epsilon_1}\frac{\partial u_1}{\partial\kappa_1} = -\frac{xu_1}{2\kappa_1} =
- \frac{xe^{\kappa_1 x}}{2\kappa_1}.
\end{equation}
We can evaluate easily the Wronskian of $u_1$ and $\frac{\partial u_1}{\partial\epsilon_1}$ by using once again equation \eqref{wrons}:
\begin{equation}
W\left(u_1,\frac{\partial u_1}{\partial\epsilon_1}\right) = -\frac{u_1^2}{2\kappa_1} = -\frac{e^{2\kappa_1 x}}{2\kappa_1}.\label{wfr}
\end{equation}
Now, replacing \eqref{wfr} into \eqref{Dv} for calculating the confluent second-order SUSY partner potential $\widetilde{V}$ of the free particle it is obtained:
\begin{equation}
\widetilde{V}=\frac{16D\kappa_1^{3}e^{2\kappa_1 x}}{(2D\kappa_1 - e^{2\kappa_1 x})^2}.\label{vfinal}
\end{equation}
Due to the `dispersion relation' ($\epsilon_1 = -\kappa_1^2, \ \kappa_1 > 0$) there is a natural restriction on the factorization energy, namely, $\epsilon_1 < 0$. Besides, in order to obtain non-singular transformations the parameter $D$ has to be restricted \cite{FS03,FS05}. Indeed, for $u_1 = e^{\kappa_1 x}$ we have that the non-singular domain is given by $D<0$, and reparametrizing as $D = -(2\kappa_1)^{-1}e^{2\kappa_1 x_0}$, $x_0\in\mathbb{R}$, we can simplify \eqref{vfinal} to obtain
\begin{equation}
\widetilde{V}=-2\kappa_1^2\text{sech}^2 [\kappa_1(x-x_0)],\label{vfinal2}
\end{equation}
which is the P\"{o}schl-Teller potential with one bound state at the energy $E_0 = \epsilon_1 = - \kappa_1^2$. It is worth to note that this result has also been obtained through first-order SUSY QM and by using the integral formulation of the confluent case \cite{FS03}. It is plausible that any non-singular SUSY transformation which departs from free particle and creates just one bound state leads precisely to a P\"oschl-Teller potential (see also \cite{FS11}).

\begin{figure}
\begin{center}
\includegraphics[scale=0.5]{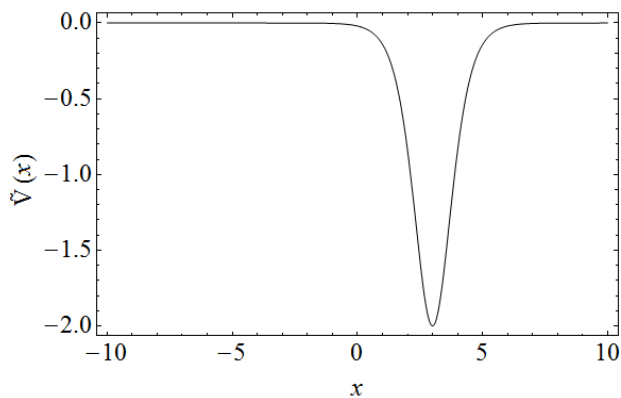}
\end{center}
\vspace{-5mm}
\caption{Confluent second-order SUSY partner potential of the free particle, obtained through Eq.~\eqref{vfinal2} for $\epsilon_1 = -1$ and $x_0 = 3$.} \label{figfree}
\end{figure}

An illustration of a confluent second-order SUSY partner potential $\widetilde V$, generated through this formalism from the free particle, is shown in Fig.~\ref{figfree}.

Note that in some previous works \cite{Mat92,Sta95}, the confluent second-order SUSY (or Darboux) transformation in this differential version was implemented for the free particle with $D=0$ and using another transformation function, namely, $u_1 = \sin [k_1(x+x_0)]$ with $\epsilon_1 = k_1^2 > 0$; however, by doing so one will deal only with singular transformations. Following the formalism of this work we have obtained a one-parameter family of non-singular potentials for each $\epsilon_1<0$.

For the free particle the integral and differential Wronskian formulae have been applied easily, since the involved integrals can be simply evaluated. Nevertheless, there are some other potentials for which the calculation of the corresponding integrals looks complicated but the differential formalism can be applied straightforwardly. We will show next an example of this situation.

\subsection{Single-gap Lam\'e potential}

The Lam\'e periodic potentials are given by \cite{Ars81,FMRS02a,FMRS02b}:
\begin{align}
V(x) &=n(n+1) m\, \text{sn}^2(x|m)\nonumber\\
	&=n(n+1)\left[\wp(x+iK(1-m))+\frac{1}{3}(m+1)\right],\label{lame}
\end{align}
where $\text{sn}(x|m)$ is a Jacobi elliptic function whose real period is $T=4K(m)$ and $\wp(x)$ is the Weierstrass elliptic function with
\begin{equation}
K(m)=\int_0^{\pi /2} \frac{d\theta}{(1-m\sin^2\theta)^{1/2}},
\end{equation}
being half the real period of $V(x)$. The potentials \eqref{lame} have $2n+1$ band edges which define $n+1$ allowed and $n+1$ forbidden bands. They belong to a class of finite-gap periodic systems where the non-linear supersymmetry plays an important role. For example, Lam\'e potentials have been used to model a non-relativistic electron in periodic electric and magnetic field configurations which produce a 1D crystal \cite{CJNP08}. In addition, these potentials admit isospectral super-extensions \cite{CJP08} and they can be used to display hidden symmetries in quantum dynamical problems, specially in soliton dynamics \cite{AS09}. Note that Lam\'e potentials are particular cases of the associated Lam\'e potentials, which have been studied previously in the context of higher-order SUSY QM \cite{FG07}.

\begin{figure}
\begin{center}
\includegraphics[scale=0.2]{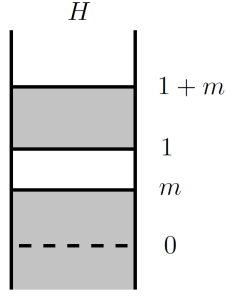}
\end{center}
\vspace{-5mm}
\caption{Spectrum of the Lam\'e potential with $n=1$. The white bands correspond to the allowed energy region, a semi-infinite $[1+m,\infty)$ and a finite one $[m,1]$. The dark region corresponds to the energy gaps, a semi-infinite $(-\infty,m)$ and a finite one $(1,1+m)$.} \label{figesplame}
\end{figure}

In this work we shall deal with the single-gap Lam\'e potential obtained with $n=1$. The spectrum for the Hamiltonian associated to this specific potential is given by:
\begin{equation}
{\rm Sp}(H) = [m,1] \cup [1+m,\infty),
\end{equation}
i.e., it is composed by a finite energy band $[m,1]$ plus a semi-infinite one $[1+m,\infty)$ (see the white region in Fig.~\ref{figesplame}). The structure of the resolvent set of $H$ is similar, namely, there is a semi-infinite energy gap $(-\infty,m)$ plus a finite one $(1,1+m)$ (observe the dark zone in Fig.~\ref{figesplame}).

As in the previous case, in order to implement the confluent second-order SUSY transformation we are going to use an appropriate seed solution $u_1$ associated to a factorization energy $\epsilon_1$ which is inside one of the energy gaps and such that $W(u_1,u_2) \neq 0 \ \forall \ x\in {\mathbb R}$. For our example this can be achieved by choosing $u_1$ as one of the two Bloch functions associated to $\epsilon_1$ \cite{FMRS02a,FMRS02b}, i.e.,
\begin{align}
u_1^{\beta}&=\frac{\sigma(\omega ')}{\sigma(\delta +\omega ')}\frac{\sigma(x+\delta +\omega ')}{\sigma(x+\omega ')}e^{-x\zeta (\delta)},\\
u_1^{1/\beta}&=\frac{\sigma(\omega ')}{\sigma(-\delta +\omega ')}\frac{\sigma(x-\delta +\omega ')}{\sigma(x+\omega ')}e^{x\zeta (\delta)},
\end{align}
where $\omega=K(m)$ and $\omega'=iK(1-m)$ are the real and imaginary half-periods of $\wp(x)$ \cite{AS64}, and $\sigma$ and $\zeta$ are the non-elliptic Weierstrass functions \cite{Cha85}.

Note that $\beta$ is defined by the relation $u_1^{\beta}(x+T)=\beta u_1^{\beta}(x)$ and then $\beta=\exp [ 2\delta\zeta(\omega)-2\omega\zeta(\delta) ]$. Besides, by expressing it as $\beta=\text{e}^{i\kappa}$, then $\kappa=2i[\omega\zeta(\delta)-\delta\zeta(\omega)]$ (up to an additive multiple of $2\pi i$) which is known as the quasi-momentum \cite{CJP08}. The displacement $\delta$ and the factorization energy $\epsilon_1$ are related by \cite{FMRS02b}:
\begin{equation}
\epsilon_1 = \frac{2}{3}(m+1)-\wp (\delta). \label{ed}
\end{equation}

In order to calculate \eqref{Dv} let us choose the first Bloch function as transformation function, namely, $u_1=u_1^{\beta}$. It is worth pointing out that we are using one Bloch state to perform the SUSY transformation, even when these states are not normalized. Nevertheless, one of the advantages of the confluent algorithm is that it does not require normalized states to perform the transformation.

We are going to calculate next its parametric derivative with respect to $\epsilon_1$, for which we will employ the following relationships between $\sigma(x)$, $\zeta(x)$, and $\wp(x)$ \cite{Cha85}:
\begin{align}
\sigma'(x)&=\sigma(x)\zeta(x),\label{d1}\\
\zeta'(x)&=-\wp(x),\\
\wp'(x)&=-\frac{\sigma(2x)}{\sigma^4(x)}.\label{d4}
\end{align}
Thus, using the chain rule and Equation \eqref{ed} it is obtained:
\begin{equation}
\frac{\partial u_1}{\partial\epsilon_1} = \frac{d\delta}{d\epsilon_1}\frac{\partial u_1}{\partial\delta} = -\left(\frac{d\wp}{d\delta}\right)^{-1}\frac{\partial u_1}{\partial\delta}.
\end{equation}
An explicit calculation produces:
\begin{equation}
\frac{\partial u_1}{\partial \delta}=[\zeta(x+\delta+\omega')-\zeta(\delta+\omega')+x\wp(\delta)]u_1,
\end{equation}
\begin{figure}[h]
\begin{center}
\includegraphics[scale=0.5]{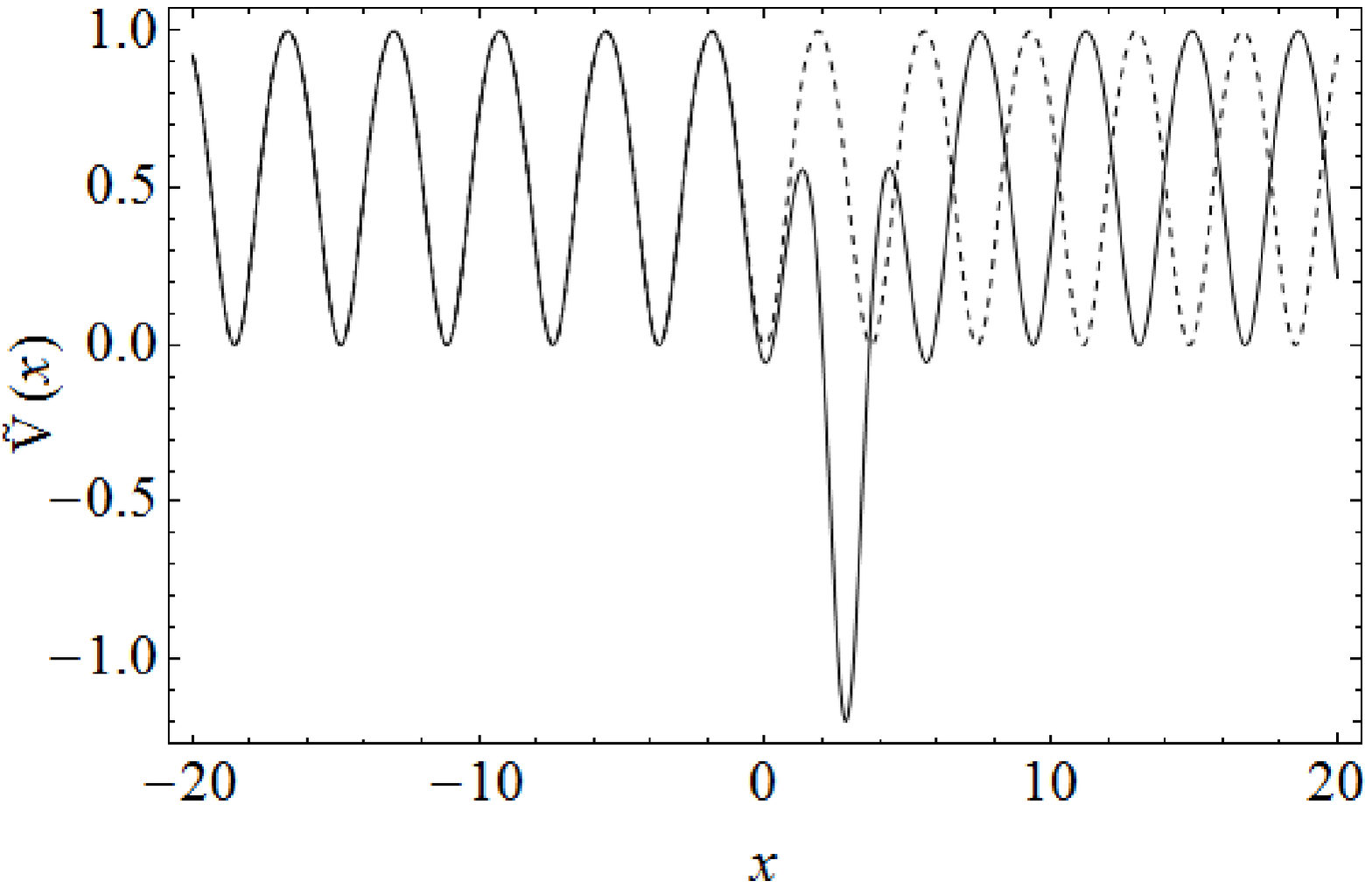}\\
\includegraphics[scale=0.47]{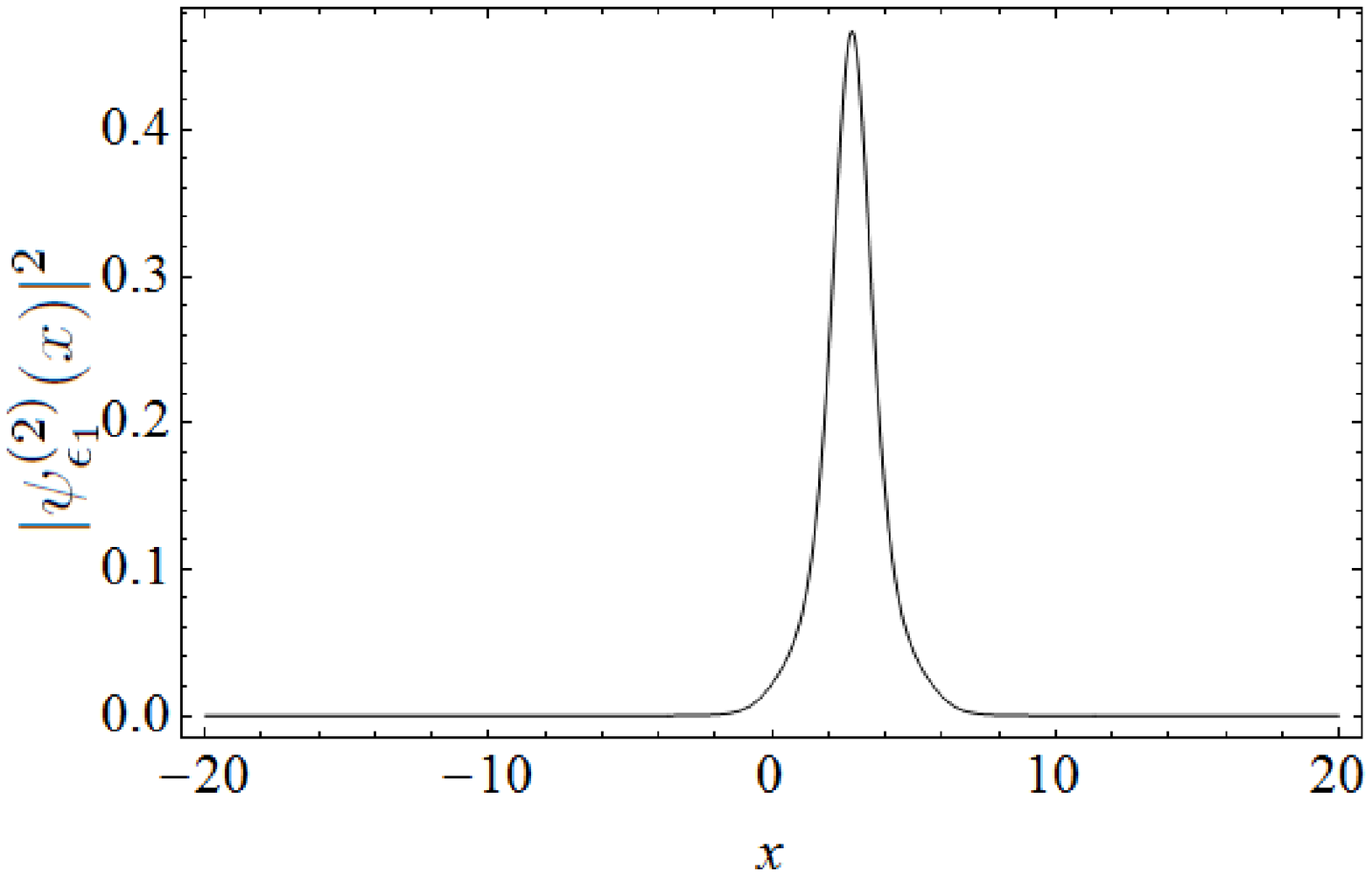}
\end{center}
\vspace{-5mm}
\caption{SUSY partner potential $\widetilde V(x)$ (top) and the probability density of its associated bound state (bottom), generated from the Lam\'e potential $V(x)$ for $n=1$. The parameters were taken as $m=0.5$, $\epsilon_1=0.1$, $x_0=0$ and $D=-45$.} \label{figlame1}
\end{figure}
and thus the Wronskian of Equation \eqref{Dv} can be obtained by using once again Equation \eqref{wrons}:
\begin{equation}
W\left(u_1,\frac{\partial u_1}{\partial\epsilon_1}\right) = \left(\frac{d\wp}{d\delta}\right)^{-1}[\wp(x+\delta+\omega')-\wp(\delta)]u_1^2\equiv f(x)u_1^{2},\label{wro}
\end{equation}
which defines the auxiliary function $f(x)$.

Finally, from equation \eqref{Dv} the new potential $\widetilde{V}$ can be calculated analytically.
\begin{equation}
\widetilde{V}=V+\frac{2+4[\zeta(x+\delta+\omega')-\zeta(x+\omega')-\zeta(\delta)](Du_1^{-2}+f)}
{(Du_1^{-2}+f)^2}.\label{Vlame}
\end{equation}

\begin{figure}[h]
\begin{center}
\includegraphics[scale=0.5]{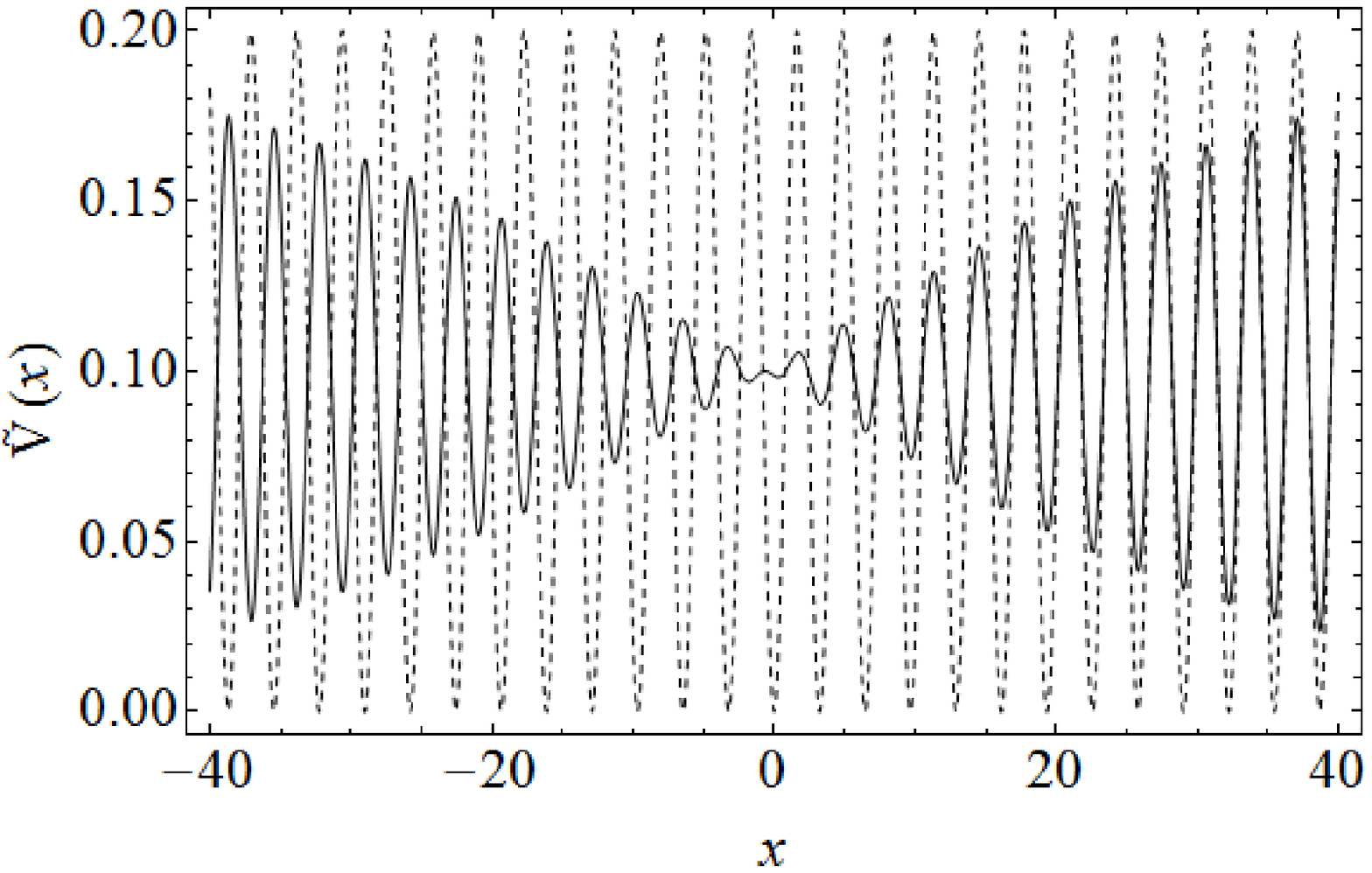}\\
\includegraphics[scale=0.5]{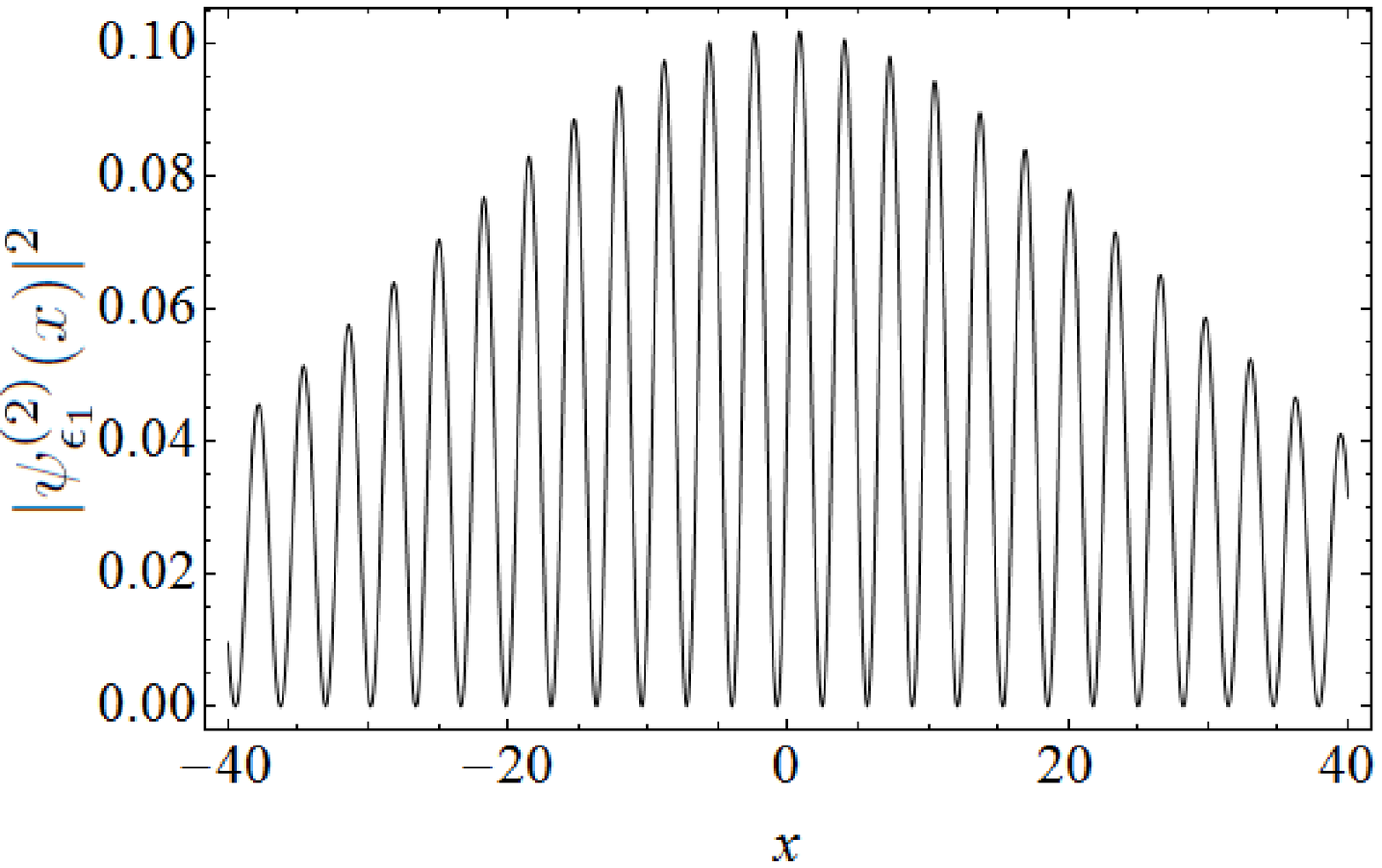}
\end{center}
\vspace{-5mm}
\caption{SUSY partner potential $\widetilde V(x)$ (top) and the probability density of its associated bound state (bottom), generated from the Lam\'e potential $V(x)$ for $n=1$. The parameters were taken as $m=0.1$, $\epsilon_1=1.05$, $x_0=0$ and $D=20$.} \label{figlame2}
\end{figure}

Two potentials obtained through this method are shown in the top of Fig.~\ref{figlame1} and Fig.~\ref{figlame2}. They correspond to two different cases, for which either the factorization energy belongs to the infinite gap or to the finite one. Note that the shape of the new potentials (continuous lines) are really different compared to the original one (dashed lines), and between them. Indeed, it can be seen that the new potentials are in general non-periodic, although they become asymptotically periodic. Note that this periodicity defect of $\widetilde V(x)$ arises due to the creation of a bound state at an energy which coincides precisely with $\epsilon_1$ (inside an initial energy gap). The width and the position of this periodicity defect in general coincides with the $x$--domain where the new bound state
\begin{equation}
\psi_{\epsilon_1}^{(2)} (x) \propto \frac{u_1}{D + W\left(u_1,\frac{\partial u_1}{\partial\epsilon_1}\right)}\label{state}
\end{equation}
has a non-trivial probability amplitude. For these two cases, the corresponding probability densities $\vert \psi_{\epsilon_1}^{(2)}(x)\vert^2$ are shown in the bottom of Fig.~\ref{figlame1} and Fig.~\ref{figlame2}.

Let us note that a similar physical situation, induced by a non-confluent second-order SUSY transformation, was found elsewhere \cite{FMRS02a,FMRS02b}. The main advantage here is that we are using just one seed solution to create a bound state inside a given energy gap. Moreover, the explicit expressions obtained from our treatment become shorter than those derived by the non-confluent algorithm. Particularly interesting is the case in which the factorization energy $\epsilon_1$ is inside the finite gap, so that a bound state is created at this position. In such a situation, if the non-periodic potential $\widetilde{V}$ is perturbed by an additional interaction, the new bound state could be used as an intermediate state to perform transitions between the finite energy band and the infinite one. Note that the new bound state of Eq.~\eqref{state} is known as {\it localized impurity state} in solid state physics (see Ch. 5 of \cite{Cal74}).

\section{Conclusions}

In this article we have introduced a differential version of the confluent second-order SUSY transformation, as an alternative to generate new exactly solvable potentials which avoids the need to evaluate some integrals arising in the formulation elaborated previously \cite{FS03,FS05}. Moreover, we have found a differential formula that generalizes the one used in soliton theory \cite{Mat92,Sta95}. The main advantage rests in the fact that families of non-singular potentials can be constructed by appropriately varying the new constant $D$ (see Eq.~\eqref{Dv}). We have successfully applied this technique to the free particle and to the single-gap Lam\'e potential. In the last case it was shown that, although the initial potentials are periodic, the SUSY generated ones are non-periodic, with a periodicity defect arising due to the creation of a bound state inside an initial energy gap. It was proposed that, under certain appropriate circumstances, this bound state could be used as an intermediate state to perform transitions from the lower energy band to the infinite one.


\section*{Acknowledgement}

The authors acknowledge the financial support of Conacyt (M\'exico) through project 152574. DB also acknowledges Conacyt PhD scholarship 219665. NFG thanks the support of SIP project 20111061, IPN-M\'exico.

\end{document}